# Geometrical and Physical Interpretations of Electronic Harmonic Oscillations in Four Space Dimensions

Kunming Xu

Environmental Science Research Center

Xiamen University, Fujian Province 361005, China

kunmingx@xmu.edu.cn

Following a previous proposition of quaternity spacetime for electronic orbitals in neon shell, this paper describes the geometrical course each electron takes as it oscillates harmonically within a certain quaternity space dimension and provides the concrete connections between geometries and trigonometric wavefunctions that observe Pythagorean theorem. By integrating four quaternity space dimensions with conventional Cartesian coordinate systems in calculus, we explain electronic motions by Maxwell's equation and general Stokes' theorem from the principles of rotation operation and space/time symmetry. Altogether with the previous reports, we have effectively established quaternity spacetime as a successful theory in elucidating the orbital shapes and motions of electrons within inert atoms such as helium and neon. We point out once again that $2p_x$, $2p_y$, and $2p_z$ orbitals have different geometrical shapes as well as orthogonal orientations, contrary to the traditional 2p orbital model.

## 1. Introduction

In a previous report [1], we introduced a fresh spacetime concept to account for a four-dimensional spherical layer such as neon shell where eight electrons are oscillating harmonically in four various space and time dimensions, obeying quaternity equation individually and simultaneously:

$$\frac{\partial^4 \Phi}{\partial t^4} = v^4 \frac{\partial^4 \Phi}{\partial l^4} \qquad (1)$$

where $\Phi$ is electronic wavefunction, $l$ is a generalized space dimension, and $t$ is a generalized time dimension in neon shell. As typical solutions to quaternity equation, we have also characterized 2s2p electrons by eight trigonometric wavefunctions:

$$\begin{pmatrix} \Phi_0 \\ \Phi_1 \\ \Phi_2 \\ \Phi_3 \\ \Phi_4 \\ \Phi_5 \\ \Phi_6 \\ \Phi_7 \end{pmatrix} = C \begin{pmatrix} \cos\Omega_0 \cos\Omega_2 - \dot{\Omega}_0 \sin\Omega_0 \cos\Omega_2 \\ -v(\sin\Omega_0 \sin\Omega_2 + \dot{\Omega}_0 \cos\Omega_0 \sin\Omega_2) \\ v^2(\cos\Omega_0 \cos\Omega_2 - \dot{\Omega}_0 \sin\Omega_0 \cos\Omega_2) \\ -v^3(\sin\Omega_0 \sin\Omega_2 + \dot{\Omega}_0 \cos\Omega_0 \sin\Omega_2) \\ v^4(\cos\Omega_0 \cos\Omega_2 + \dot{\Omega}_0 \sin\Omega_0 \cos\Omega_2) \\ -v^3(\sin\Omega_0 \sin\Omega_2 - \dot{\Omega}_0 \cos\Omega_0 \sin\Omega_2) \\ v^2(\cos\Omega_0 \cos\Omega_2 + \dot{\Omega}_0 \sin\Omega_0 \cos\Omega_2) \\ -v(\sin\Omega_0 \sin\Omega_2 - \dot{\Omega}_0 \cos\Omega_0 \sin\Omega_2) \end{pmatrix} \quad (2)$$

$$\dot{\Omega}_0 = -\frac{\partial \Omega_0}{\partial t} \quad (3)$$

where $C$ is a constant, $v$ denotes a generalized velocity dimension, $\dot{\Omega}_0$ is a complex number notation and a gateway function connecting both 1s and 2s2p layers, and $\Omega_0$ and $\Omega_2$ are time and space radian angles respectively, relating to two curvilinear vectors of 1s electrons that observe rotation relation:

$$-\frac{\partial \Omega_0}{\partial t} = \frac{\partial \Omega_2}{\partial l} \quad (4)$$

Between every two adjacent electrons in neon shell, there is always a rotation relation distance, i.e., a differential operation with respect to a minus time dimension accompanied by an integral operation over a space dimension. For instance, the relationship between wavefunctions $\Phi_0$ and $\Phi_1$ can be characterized by

$$\Phi_1 = \int -\frac{\partial \Phi_0}{\partial t} dl \quad (5)$$

or

$$-\frac{\partial \Phi_0}{\partial t} = \frac{\partial \Phi_1}{\partial l} \quad (6)$$

which can be derived by applying relationship (4) to wavefunctions in equation (2). For brevity, we shall only consider the relationship of their first terms because the first term can be treated as a dynamic pointer pointing towards the head of the whole electronic curvilinear vector.

$$-\frac{\partial \Phi_{01}}{\partial t} = \frac{\partial \Phi_{11}}{\partial l} \quad (7)$$

where $\Phi_{01}$ and $\Phi_{11}$ refer to the first term of $\Phi_0$ and $\Phi_1$ in equation (2) respectively. Equation (7) indicates that the rate of time component reduction from a full time dimension to vanishing is equal to the rate of space component change from a full space dimension to vanishing, which can be equivalently expressed in trigonometry as:

$$\cos\Omega_0 = \sin\Omega_2 \quad (8)$$

Thus radians $\Omega_0$ and $\Omega_2$ constitute two acute angles of a right triangle so that equation (8)

is an alternative expression of Pythagorean theorem. By the way, since both complementary angles represent time and space radians, respectively, they must coexist as a product in the wavefunction, i.e., one cannot substitute the other.

The forgoing deduction has adopted the following new interpretation of calculus on trigonometric functions:

**Definition 1.** For a trigonometric function such as $y = \cos\theta$, a differential operation of $y$ with respect to radian $\theta$ means increasing $\theta$ variable a displacement of $\pi/2$ in the function.

**Definition 2.** For a trigonometric function such as $z = \cos\vartheta$, an integral operation of $z$ over $\vartheta$ means reducing $\vartheta$ variable a displacement of $\pi/2$ in the function.

The correctness of both definitions can be easily proved by the property of cosine and sine functions in calculus. The difference between conventional differentiation concept and the above radian angle rotation definitions is that the latter traverses the full range of $\pi/2$ radian whereas the former only catches the terminal state at a certain radian value. The latter is actually the dynamic implementation of the former. Thus, in a more general manner, the process of a differential operation on a wavefunction can be expressed as a cosine or sine function with a changing radian variable, which leads to two theorems with regard to dynamic differentiation:

**Theorem 1.** For a trigonometric wavefunction $y$, the derivative term $-\partial y/\partial t$ can be expressed as the changing rate of variable $y$ from a full time dimension $t$ to vanish following a cosine rule such as

$$-\frac{\partial y}{\partial t} = A\cos\theta \tag{9}$$

where $A$ is an orthogonal quantity to $\cos\theta$, $t$ is a fix time dimension, and the sinuously changing range of $y$ (t, 0) value corresponds to radian range $\theta$ (0, $\pi/2$) in the dynamic process.

**Theorem 2.** For a trigonometric wavefunction $z$, the derivative term $\partial z/\partial l$ can be expressed as the changing rate of variable $z$ from a full space dimension $l$ to vanish following a sine rule such as:

$$\frac{\partial z}{\partial l} = A\sin\vartheta \tag{10}$$

where $A$ is an orthogonal quantity to $\sin\vartheta$, $l$ is a fix space dimension, and the sinuously changing range of $z$ (l, 0) value corresponds to radian range $\vartheta$ ($\pi/2$, 0) in the dynamic process.

By interpreting differential operations in terms of trigonometric functions, we have provided the implementation for dynamic differential processes. Conventional differential operation is only concerned with the final result of the operation, ignoring the intermediary course or process. However, the result is only instantaneous locating at a discrete point of $\theta$ (such as $\theta = \pi/2$) along a continuous curve while the process is perennial covering the full range of $\theta\,(0,\ \pi/2)$ variable in the smooth curve. This sheds light on the relationship between the discrete variable and the continuous variable, i.e., the discrete result is only a special point on the continuous curve. Thus the dynamic process of differential operation upon a wavefunction is richer in physical meaning than its mere final result. Given these considerations, we shall examine the process of wavefunction transformations from one state to another via their dynamic pathways in trigonometry and geometry in the following section.

## 2. Pythagorean theorem in quaternity space

The electron octet as were listed in equation (2) form four pairs of conjugated complex functions, each pair with four velocity dimensions apart: $\Phi_0$ and $\Phi_4$ are a pair of 2s electrons; $\Phi_1$ and $\Phi_5$ are a pair of $2p_x$ electrons; $\Phi_2$ and $\Phi_6$ are a pair of $2p_y$ electrons; and $\Phi_3$ and $\Phi_7$ are a pair of $2p_z$ electrons (Figure 1). Because of their various orbital types, electronic rotations from one state to another make four distinctive spatial courses that obey Pythagorean theorem in four unique right triangles. In the following subsections, we shall consider the geometry of four electronic transformations $\Phi_0 \mapsto \Phi_1 \mapsto \Phi_2 \mapsto \Phi_3 \mapsto \Phi_4$ in details. On the opposite sides of quaternity coordinates, another four electronic rotation steps $\Phi_4 \mapsto \Phi_5 \mapsto \Phi_6 \mapsto \Phi_7 \mapsto \Phi_0$ observe similar transformation principle but with regard to space and time dimension switches.

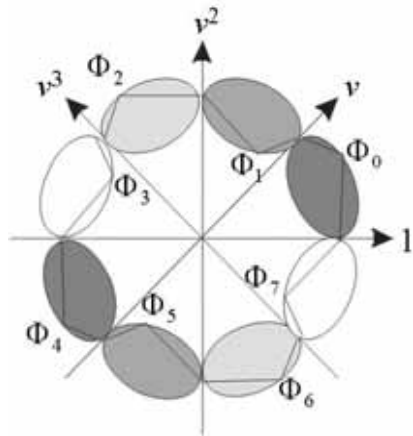

Figure 1. Transformation pathways of four conjugated electronic pairs that form four closed circles on quaternity coordinates where each pair orbit along both semicircular arcs of the same shaded circle in the abstract.

**2.1 One-dimensional harmonic oscillation**

Figure 2(a) indicates the state change of a 2s electron to a $2p_x$ type along straight line OB. The 2s electron, which initially permeates the entire sphere as dimensionless cloud, gradually contracts its activity sphere while its spherical center moves towards point B, which may represent a solid charge. The transformation corresponds to wavefunction evolution from $\cos\Omega_0 \cos\Omega_2$ to $-\dot{\Omega}_0 \sin\Omega_0 \cos\Omega_2$ in time and from $-\dot{\Omega}_0 \sin\Omega_0 \cos\Omega_2$ to $-v\sin\Omega_0 \sin\Omega_2$ in space as was indicated by equation (5). The initial term $\cos\Omega_0 \cos\Omega_2$ can be regarded as a four-dimensional time with zero dimensional space. Time dimension reduction is undergoing with space dimension increment as the electron transforms from misty cloud to a real particle. With the increase of $\Omega_2$ angle from 0 to π/2, the temporal radius of sphere O reduces from OB to EB. At a specific moment, center O moves to center E, which is the vertical projection of point A along a smooth semicircular arc OAB. In other words, spherical center E is traveling along straight line OB but follows harmonic oscillation principle as if it travels uniformly along semicircular arc OAB.

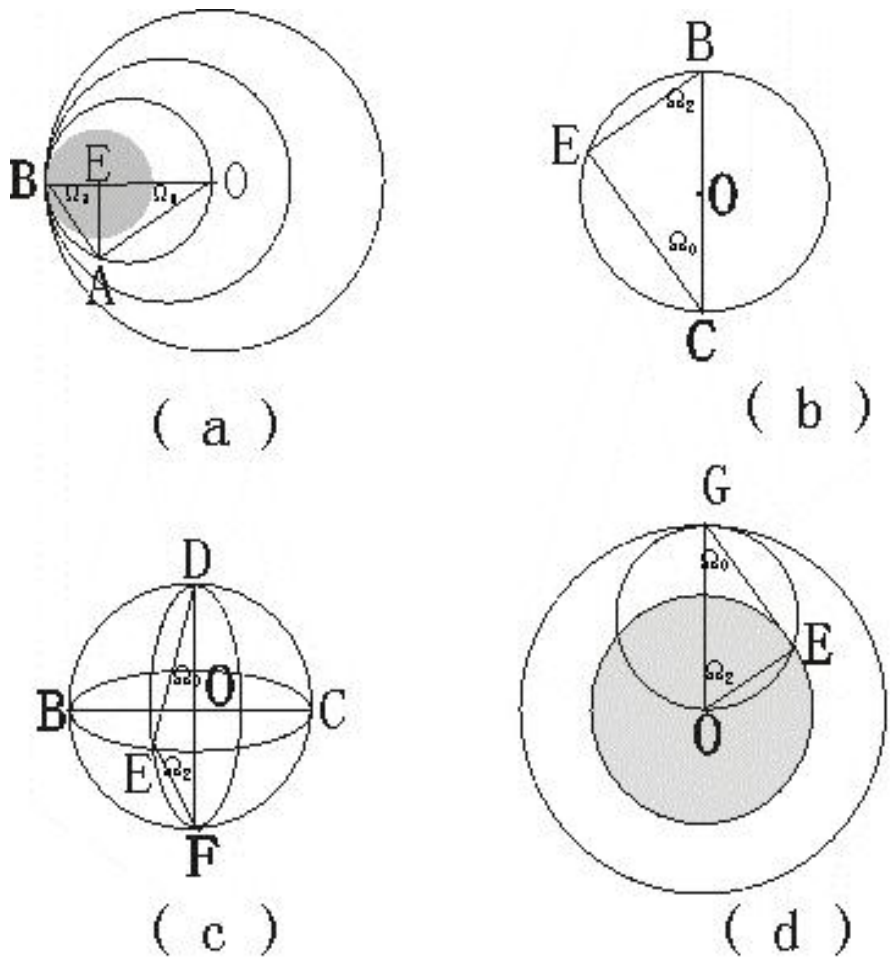

Figure 2 Electronic transformations track along semicircular arcs of various space dimensions in four harmonic oscillations: (a) $2p_x$-oribtal, (b) $2p_y$-orbital, (c) $2p_z$-orbital, and (d) 2s-orbital.

Sphere O is the maximum four-dimensional time sphere that is ever attained by 2s electron with a wavefunction of $\cos\Omega_0 \cos\Omega_2$ whereas point A is $-v\sin\Omega_0 \sin\Omega_2$ term representing a 2p$_x$ electronic state. In right triangle OAB, side AB is the time component of the electron, corresponding to the product of hypotenuse OB and $\cos\Omega_2$, projecting into radius EB, and decreasing, whereas side AO is its space displacement from its initial state, i.e., particle size, corresponding to the product of hypotenuse OB and $\sin\Omega_2$ factor, which is increasing. The semicircular orbit OAB is the pathway for the electron to cover the abstract distance of diameter OB, the full 1D space element, or the final spatial displacement of the electron from its initial 2s state. At point B, the electron has possessed one-dimensional space OB, and has three-dimensional time accordingly. As the electron traverses along this 1D space dimension, we shall call this dynamic process a 2p$_x$ orbital.

**2.2 Two-dimensional harmonic oscillation**

As shown in Figure 5(b), a 2p$_y$ electron *E* is traveling along a semicircular arc from points B to C as radian angle $\Omega_2$ increases from π/2 to π (i.e., decreases from π/2 to 0 in right triangle BEC). The electron may manifest as magnetic flux during the process. In right triangle BEC, point B denotes $-v\dot\Omega_0 \cos\Omega_0 \sin\Omega_2$ term spatially and point C denotes $v^2\dot\Omega_0 \cos\Omega_0 \cos\Omega_2$ term spatially. At a specific moment along the path, side BE indicates space component of 2p$_y$ electron while side CE is time component of the electron. Space dimension is increasing while time dimension is decreasing. Diameter BC corresponds to a full 2D space element for the electron to attain, and semicircular arc BEC is the pathway to traverse that dimension. As the electron at B tracks along the semicircular arc to C, it draws the shape of a 2p$_y$ orbital as a semicircular ring on the plane.

**2.3 Three-dimensional harmonic oscillation**

Figure 5(c) explains the motion of a 2p$_z$-electron. It is the revolution of semicircular arc BEC around the axis BC. The position of whole arc BEC is determined by angle $\Omega_2$ in right triangle DEF. With the increase of $\Omega_2$ angle from π to 3π/2 (i.e., from 0 to π/2 in right triangle DEF), arc BEC sweeps a hemispherical surface, which gives the shape of a full 2p$_z$ orbital. This initial arc position BDC denotes $-v^2\dot\Omega_0 \sin\Omega_0 \cos\Omega_2$ spatially whereas final arc BFC denotes $-v^3 \sin\Omega_0 \sin\Omega_2$ spatially. At a specific moment along p$_z$-orbital, the interval between arcs BDC and BEC measured by side DE in right triangle DEF represents space component of the electron whereas chord FE denote its time component. Space increases while time dwindles. A full 2p$_z$ orbital has wrapped the 3D hemispherical surface represented by diametrical chord DF. Its geometrical shape is a hollow hemispherical surface such as BDCFE as was shown in Figure 2(c).

**2.4 Four-dimensional harmonic oscillation**

Figure 5(d) indicates electronic transformation from $2p_z$ to 2s types, which involves the expansion of a spherical surface, in which the radius OE, the maximum radius OG, and tangential line GE form a right triangle. The radial expands via the course of arc OEG. Point O denotes $-v^3 \dot{\Omega}_0 \cos \Omega_0 \sin \Omega_2$ whereas point G denotes $v^4 \dot{\Omega}_0 \sin \Omega_0 \cos \Omega_2$ function in spacetime. With the increase of $\Omega_2$ angle from $3\pi/3$ to $2\pi$ (i.e., decrease from $\pi/2$ to 0 in right triangle OEG), side OE is space component of 2s electron, which is increasing gradually, whereas side GE is time component of the electron, which diminishes in the meanwhile. Arc OEG is the pathway to traverse the 4D space element represented by diametrical chord OG.

As the whole spherical surface expands along the radial direction, the electron disperses over the entire sphere so that the geometrical shape of 2s electron is a continuous solid sphere of four dimensions in space. A full 2s orbital, whose time dimension becomes zero rendering the whole spherical space instantaneous, proceeds to further oscillation cycles via $\Phi_4 \mapsto \Phi_5 \mapsto \Phi_6 \mapsto \Phi_7 \mapsto \Phi_0$ steps and finally returns to original state $\Phi_0$.

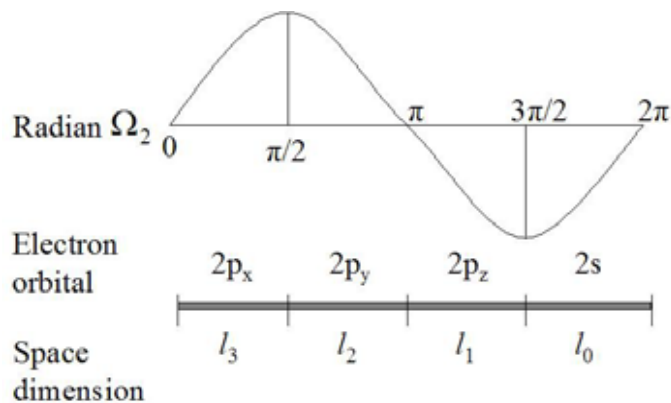

Figure 3 Harmonic oscillations of electrons in one, two, three, and four space dimensions in correspondence with radian angle $\Omega_2$ revolution in a $2\pi$ cycle.

The forgoing description demonstrated the same principle of right triangles, i.e., Pythagorean theorem as was expressed by equation (8). A 2p-orbital is one-dimensionally oriented in space; a $2p_y$-oribital moves on a two-dimensional planar ring; a $2p_z$-electron sweeps around a three-dimensional hemispherical surface; and a 2s-orbital permeates over the four-dimensional sphere. These provide the visual connections between geometries and trigonometric wavefunctions as were listed in equation (2). Four distinctive diametrical chords OB, BC, DF, and OG in Figure 2(a), 2(b), 2(c), and 2(d), respectively, represent four space dimensions for an electron to traverse in $\Phi_0 \mapsto \Phi_1 \mapsto \Phi_2 \mapsto \Phi_3 \mapsto \Phi_4$ processes respectively via their subtending semicircular arcs. We shall label these quaternity space dimensions as $l_3$, $l_2$, $l_1$, and $l_0$ that correspond to radian $\Omega_2$ revolution in a full period (Figure 3), their spatial orientations being 1D, 2D, 3D, and 4D, respectively. We may generalize $l_3$, $l_2$, $l_1$, or $l_0$ dimension as $l$, which refers to the outstanding space dimension of concern or the most immediate dimension in the wavefunction of concerned.

### 3. Quaternity space in Cartesian coordinates

So far we have identified four distinct space dimensions $l_3$, $l_2$, $l_1$, and $l_0$ in spherical layer of neon shell, but how these dimension elements are related to Cartesian coordinates in X, Y, and Z orientations? First of all, consider $l_3$ dimension in Figure 2(a), if we set up a one-dimensional X axis along OB direction with the origin at point O, then for the electron to traverse $l_3$ dimension is equivalent to moving along X-axis, which gives the conformity of $l_3$ and X measures for $2p_x$ transformation in one-dimensional harmonic oscillation:

$$\frac{\partial \Phi_1}{\partial l_3} = \frac{\partial \Phi_1}{\partial x} \tag{11}$$

Secondly, in Figure 2(b) for $2p_y$-orbital transformation, we may add Y-axis to X-axis to form a two-dimensional Cartesian plane as shown in Figure 4. Chord BE is the space component of the electron in quaternity, which form a hypotenuse in right triangle BEP. Following the rule of vector addition, we have

$$\overrightarrow{BE} = \overrightarrow{BP} + \overrightarrow{PE} \tag{12}$$

where vector $\overrightarrow{BE}$ can be expressed as $\partial \Phi_2 / \partial l_2$ in quaternity space in the dynamic transformation, vector $\overrightarrow{PE}$ can be expressed as $\partial \Phi_{2y} / \partial x$ in the same dynamic sense, and $\overrightarrow{BP}$ is equal to ($\overrightarrow{OB} - \overrightarrow{OP}$), which can be expressed differentially as $-\partial \Phi_{2x} / \partial y$ because $\overrightarrow{OB}$ is a constant radius. Thus, equation (12) becomes:

$$\frac{\partial \Phi_2}{\partial l_2} = \frac{\partial \Phi_{2y}}{\partial x} - \frac{\partial \Phi_{2x}}{\partial y} \tag{13}$$

or in two-dimensional Curl operator expression as

$$\frac{\partial \Phi_2}{\partial l_2} = \nabla \times \Phi_2 \tag{14}$$

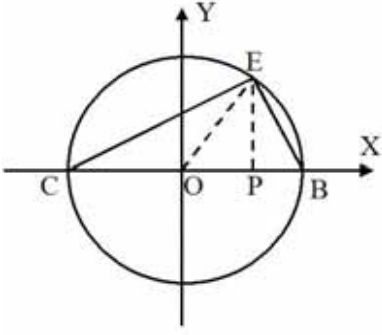

Figure 4 Schematic diagram for establishing vector relationship between quaternity space and Cartesian coordinates.

Furthermore, with regard to $2p_z$ electronic transformation on spherical surface as was shown in Figure 2(c), we may add Z-axis to X-Y axes to form a three-dimensional coordinates system. By expressing the same spherical surface vector in terms of quaternity space and of Cartesian coordinates, we obtain

$$\frac{\partial \Phi_3}{\partial l_1} = (\frac{\partial \Phi_{3z}}{\partial y} - \frac{\partial \Phi_{3y}}{\partial z})\mathbf{i} + (\frac{\partial \Phi_{3x}}{\partial z} - \frac{\partial \Phi_{3z}}{\partial x})\mathbf{j} + (\frac{\partial \Phi_{3y}}{\partial x} - \frac{\partial \Phi_{3x}}{\partial y})\mathbf{k} \qquad (15)$$

where $\mathbf{i}$, $\mathbf{j}$, and $\mathbf{k}$ are unit vectors along X, Y, and Z directions respectively. Because $l_1$ denotes a spherical surface dimension, the addition of three vectors, $\mathbf{i}$, $\mathbf{j}$, and $\mathbf{k}$, amounts to a vector of constant length that equals the spherical surface radius. With three-dimensional Curl operator, equation (15) can be expressed compactly by

$$\frac{\partial \Phi_3}{\partial l_1} = \nabla \times \Phi_3 \qquad (16)$$

In light of this, we may express the principle of rotation operation

$$-\frac{\partial \Phi_2}{\partial t} = \frac{\partial \Phi_3}{\partial l_1} \qquad (17)$$

as

$$-\frac{\partial \Phi_2}{\partial t} = \nabla \times \Phi_3 \qquad (18)$$

which is in the same form as Maxwell's equation that relates electric field strength $E$ to magnetic field strength $B$.

$$-\frac{\partial B}{\partial t} = \nabla \times E \qquad (19)$$

This means that the relationship between $2p_y$ and $2p_z$ orbitals observes electromagnetism strictly. The relationship between $2p_x$ and $2p_y$ orbitals has the same property.

Finally, from Figure 2(d) for 2s-orbital transformation, we can find the divergence operator as a differential reflection of Pythagorean theorem for three perpendicular

orientations:

$$\overrightarrow{OE} = X\mathbf{i} + Y\mathbf{j} + Z\mathbf{k} \tag{20}$$

whence

$$\frac{\partial \Phi_4}{\partial l_0} = (\frac{\partial}{\partial x} + \frac{\partial}{\partial y} + \frac{\partial}{\partial z})\Phi_4 \tag{21}$$

or in div operator expression as

$$\frac{\partial \Phi_4}{\partial l_0} = \nabla \cdot \Phi_4 \tag{22}$$

In light of this, we may express the principle of rotation operation

$$-\frac{\partial \Phi_3}{\partial t} = \frac{\partial \Phi_4}{\partial l_0} \tag{23}$$

as

$$-\frac{\partial \Phi_3}{\partial t} = \nabla \cdot \Phi_4 \tag{24}$$

which express the general continuity relation between current density *J* and the charge density $\rho$ at a point:

$$-\frac{\partial \rho}{\partial t} = \nabla \cdot J \tag{25}$$

Thus we have developed the relationships between four dimensions in quaternity space and traditional Cartesian coordinates in differential form. These integrations are very important for accurately understanding the concept of quaternity space and lay the foundation for physical integration of electronic harmonic oscillations with electromagnetism. One may interpret the principle of rotation operation in diverse physical manners.

## 4. Equilibrium of electrons by Stokes' theorem

We have described electronic motions by trigonometry and geometry as smooth and continuous spacetime evolution processes. Each pair of $2p_x^2$, $2p_y^2$, $2p_z^2$, and $2s^2$ electrons oscillate harmonically along a circle of one, two, three, and four spatial dimensions respectively. But how electrons synchronize their motions in various orbital states? We have shown that electronic rotation complies with electromagnetic laws in differential form. Here we shall further investigate the principle of rotation operation in integral form.

### 4.1 Principle of rotation operation and spacetime symmetry

Every two adjacent electrons obey the principle of rotation operation such as equation (6) where space dimension *l* refers to $l_3$ that is one-dimensional in correspondence with X-axis as was formulated by equation (11). Equation (6) describes the relationship between a 2s electron and a 2p$_x$ electron in differential form.

By the principle of space and time symmetry, differentiations upon a wavefunction respect with to space dimension and to time dimension must be equal after balancing their resultant dimensions:

$$\frac{\partial \Phi_0}{\partial t} = v_1 \frac{\partial \Phi_0}{\partial l_0} \tag{26}$$

where $v_1$ is a velocity serving as a dimension compensator. Combining equations (6) and (26) produces

$$v_1 \frac{\partial \Phi_0}{\partial l_0} = -\frac{\partial \Phi_1}{\partial l_3} \tag{27}$$

which means that the rate of a 2s spherical contraction is proportional to the rate of a 2p$_x$ displacement from the center nucleus during their full cycle oscillations (Figure 2a). The same principles can be expressed in integral form as follows.

### 4.2 Green's theorem

In integral form, equation (6) can be expressed as

$$\int -\Phi_0 dl_3 = \int \Phi_1 dt \tag{28}$$

By the symmetry of space and time components, integrations of $\Phi_1$ over time and over space must be equal after balancing their resultant dimensions.

$$v_2 \int \Phi_1 dt = \int \Phi_1 dl_2 \tag{29}$$

Combining equations (28) and (29) yields

$$-v_2 \int \Phi_0 dl_3 = \int \Phi_1 dl_2 \tag{30}$$

or in X-Y coordinates of

$$-v_2 \int \Phi_0 dx = \int \Phi_{1x} dx + \Phi_{1y} dy \tag{31}$$

which means that the integral of $\Phi_1$ over semicircle BEC is counterbalanced by the integral of $\Phi_0$ over diameter COB (Figure 5).

According to Green's theorem, line integral of $\Phi_1$ over the entire circle BECPB represents the circular orbital motion of both 2p$_y$ electrons, which equals the area integral over the enclosed circular region:

$$\oint_C \Phi_{1x} dx + \Phi_{1y} dy = \iint_R (\frac{\partial \Phi_{1y}}{\partial x} - \frac{\partial \Phi_{1x}}{\partial y}) dx dy \tag{32}$$

where $C$ refers to the periphery of the circle and $R$ refers to the enclosed area. Hence we obtain the integral expression of quaternity space in Cartesian coordinates with a curl operator:

$$\oint_C \Phi_1 dl_2 = \iint_R (\nabla \times \Phi_1) dxdy \tag{33}$$

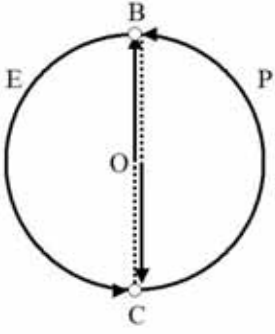

Figure 5 Oscillation pathways of two $2p_y$ (arcs BEC and CPB) and two $2p_x$ (diameters COB and BOC) where the dotted lines are parts of $2p_x$ orbitals though not in the spherical center track of the orbitals.

**4.3 Stokes' theorem**

By the principle of rotation operation and space and time symmetry, we may obtain a similar relationship between $2p_y$ and $2p_z$ electronic orbitals to equation (30).

$$-v_3 \int \Phi_1 dl_2 = \int \Phi_2 dl_1 \tag{34}$$

where $v_3$ is a velocity dimension. The right-hand side of the equation refers to the area integral of $\Phi_2$ over the smooth spherical surface and the left-hand side is the line integral of $\Phi_1$ over the spherical surface boundary (Figure 6). In Cartesian coordinates, the right-hand side of equation (34) can be expressed as

$$\int \Phi_2 dl_1 = \iint_H \left(\frac{\partial \Phi_{2z}}{\partial y} - \frac{\partial \Phi_{2y}}{\partial z}\right) dydz + \left(\frac{\partial \Phi_{2x}}{\partial z} - \frac{\partial \Phi_{2z}}{\partial x}\right) dzdx + \left(\frac{\partial \Phi_{2y}}{\partial x} - \frac{\partial \Phi_{2x}}{\partial y}\right) dxdy \tag{35}$$

or written compactly as

$$\int \Phi_2 dl_1 = \iint_H (\nabla \times \Phi_2) dA \tag{36}$$

where $H$ refers to the hemispherical surface of a $2p_z$ orbital.

From equations (34) and (36) and taking both $2p_y$ and both $2p_z$ electrons into considerations, we get

$$-v_3 \oint_C \Phi_{1x} dx + \Phi_{1y} dy = \oiint_H (\nabla \times \Phi_2) dA \tag{37}$$

This equation is in the form of Stokes' theorem that the line integral along the boundary circle of both $2p_y$ orbitals counterbalances the area integral over spherical surface of both $2p_z$ orbitals.

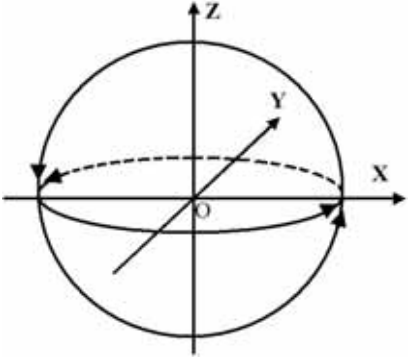

Figure 6 Two semicircular arcs in X-Y plane form the line boundary of the upper and lower hemispherical surfaces.

**4.4 Gauss' theorem**

Gauss' theorem is commonly known as divergence theorem. Let $V$ be the spherical region of 2s electron in space with spherical surface boundary $H$, then the volume integral of the divergence $\nabla \cdot \Phi_3$ of $\Phi_3$ over $V$ and the surface integral of $\Phi_3$ over the boundary $H$ of $V$ are related by

$$\int_V (\nabla \cdot \Phi_3) dV = \int_H \Phi_3 dH \tag{38}$$

On the other hand, in quaternity space we may derive a relationship between $2p_z$ and 2s electrons:

$$-v_4 \int \Phi_2 dl_1 = \int \Phi_3 dl_0 \tag{39}$$

which parallels equations (30) and (34). We realize that the right-hand side of equation (39) is equivalent to the left-hand side of equation (38) concerning the volume integral over the sphere.

$$\int \Phi_3 dl_0 = \iiint (\nabla \cdot \Phi_3) dxdydz \tag{40}$$

whence

$$-v_4 \int \Phi_2 dl_1 = \iiint_V (\nabla \cdot \Phi_3) dxdydz \tag{41}$$

or

$$-v_4 \iint_H (\nabla \times \Phi_2) dA = \iiint_V (\nabla \cdot \Phi_3) dxdydz \tag{42}$$

which indicates that the density changes of 2s electrons within the region of space $V$ is counterbalanced by the flux into or away from that region through its spherical surface boundary of $2p_z$ electrons.

The forgoing discussion has used $v_1$, $v_2$, $v_3$, and $v_4$ for referring to various velocity dimensions corresponding to $l_3$, $l_2$, $l_1$, and $l_0$ space dimensions with $t_1$, $t_2$, $t_3$, and $t_4$ time dimensions (which we did not distinguish above), respectively. The generalized $v$ refers to the proper velocity dimension of concerned. We may further write equations (31), (37) and (42) in a generalized form:

$$-v\int d\Phi_i = \iint d\Phi_{i+1}; i = 0,1,2,3 \tag{43}$$

where $\Phi_i$ and $\Phi_{i+1}$ are two adjacent electronic orbitals that are coupled together through general Stokes' theorem. Thus each electronic orbital is in dynamic equilibria with its adjacent ones.

We have found that the principle of rotation operation is closely connected with general Stokes' theorem. Since Green's theorem, Stokes' theorem, and Gauss' theorem have broad physical implications in dynamics, fluid dynamics, electromagnetism, heat conduction, thermodynamics, etc., one may explain electronic behavior in diverse physical manners. This discovery opens up prospects for unifying classic mechanics with quantum mechanics. For example, $\oint \Phi_1 dl_2$ is an integral about a closed circle and can be regarded as the circulation of the velocity field in fluid dynamics, which corresponds to Z-component of the angular momentum of the electron described by colatitude quantum number in quantum mechanics.

## 5. Summary

We have described electronic orbitals in neon shell in reasonable geometrical details beyond quantum mechanics. Electrons are oscillating simultaneously and individually in harmonic ways in four various dimensions: a $2p_x$ electron is one-dimensionally oriented; a $2p_y$ is tracking along two-dimensional semicircular arc; a 2p electron assembles on three-dimensional hemispherical surface; and a 2s electron permeates over the entire atomic sphere that is four-dimensional in quaternity space. These four space dimensions are four diameters that electrons are traversing via their corresponding semicircular arcs during harmonic oscillations. It is interesting that electrons track along the circular arcs, but their space and time components at any moment are measured by their subtended chords. The space and time components of the electron as two sides and the diametrical dimension as a hypotenuse always form a right triangle so that Pythagorean theorem indeed governs electronic motions in any cases.

We have identified four space dimensions in neon shell as $l_3$, $l_2$, $l_1$, and $l_0$ and established their mathematical connections with conventional Cartesian coordinates in differential forms (see equations 11, 13, 15, and 22) as well as in integral forms. We summarize the integral definitions of four qauternity space dimensions under the context of electronic wavefunctions in neon shell.

$$\begin{cases} \int \Phi_0 dl_3 = \int \Phi_0 dx \\ \int \Phi_1 dl_2 = \int \Phi_{1x} dx + \Phi_{1y} dy \\ \int \Phi_2 dl_1 = \iint_H (\frac{\partial \Phi_{2z}}{\partial y} - \frac{\partial \Phi_{2y}}{\partial z}) dydz + (\frac{\partial \Phi_{2x}}{\partial z} - \frac{\partial \Phi_{2z}}{\partial x}) dzdx + (\frac{\partial \Phi_{2y}}{\partial x} - \frac{\partial \Phi_{2x}}{\partial y}) dxdy \\ \int \Phi_3 dl_0 = \iiint (\frac{\partial}{\partial x} + \frac{\partial}{\partial y} + \frac{\partial}{\partial z}) \Phi_3 dxdydz \end{cases} \quad (44)$$

These equations effectively integrated quaternity space definition with Cartesian coordinate system and made it easy to understand from the conventional perspective. It was striking that Maxwell's equation, general Stokes' theorem, and Pythagorean theorem boiled down to the same law as the principle of rotation operation that we have proposed for electronic motions since our first report [2].

Each electronic orbital is a continuous process driven by a radian angle rotation under the new definition of dynamic differentiation and all electronic processes are contiguous in spacetime. Each electron is in equilibria with its adjacent ones. Electronic interaction obeys general Stokes' theorem that connotes rich physical significances. One may interpret electronic behavior as many ways as general Stokes' theorem in physical applications. This original article is a further exploration beyond the previous reports. Readers are encouraged to consult the previous article [1] and monograph [2] should any conceptual questions arise.